\documentclass[12pt,preprint,showpacs,preprintnumbers,amsmath,amssymb,prl]{revtex4}

\usepackage{graphicx}

\def\d{{\rm d}}

\begin{document}

\preprint{APS/123-QED}

\title{Theory of localized synfire chain}
\author{Kosuke Hamaguchi}
\affiliation{%
RIKEN BSI, 2-1 Hirosawa Wako, Saitama, JAPAN
}%
\author{Masato Okada}%
\affiliation{%
RIKEN BSI, 2-1 Hirosawa Wako, Saitama, JAPAN
}%
\affiliation{%
Department of Complexity Science and Engineering, University of Tokyo,  Kashiwanoha 5-1-5, Kashiwa, Chiba, 277-8561, JAPAN}%
\affiliation{%
PRESTO JST, JAPAN}%
\author{Kazuyuki Aihara}
\affiliation{%
IIS, University of Tokyo and ERATO Aihara Complexity Modeling Project JST\\
Komaba 4-6-1, Meguro, Tokyo 153-8505, JAPAN
}%

\date{\today}

\begin{abstract}
Neuron is a noisy information processing unit and conventional view is that information in the cortex is carried on the rate of neurons spike emission. More recent studies on the activity propagation through the homogeneous network have demonstrated that signals can be transmitted with millisecond fidelity; this model is called the Synfire chain and suggests the possibility of the spatio-temporal coding. However, the more biologically realistic, structured feedforward network generates spatially distributed inputs. It results in the difference of spike timing. This poses a question on how the spatial structure of a network effect the stability of spatio-temporal spike patterns, and the speed of a spike packet propagation.
By formulating the Fokker-Planck equation for the feedforwardly coupled network with Mexican-Hat type connectivity, we show the stability of localized spike packet and existence of Multi-stable phase where both uniform and localized spike packets are stable depending on the initial input structure. The Multi-stable phase enables us to show that a spike pattern, or the information of its own, determines the propagation speed.
\end{abstract}

\maketitle
Neuronal synchrony observed in the cortex is thought to play functional roles like binding features during cognitive processes \cite{Nature:Gray89}.
The mechanism of synchrony has theoretically been studied in several neural network models, e.g., pulse coupled oscillators \cite{Mirollo90,Kuramoto91,Vreeswijk96}, chaotic oscillators \cite{PRL:Rosenblum1996}, and feedforwardly connected networks \cite{Abeles1991,Diesmann99,Cateau2001,Kistler2002,JNeurosci:Litvak2003}. 
In general, experimental verification is required for the development of science, 
but many theoretical works on the level of neuronal circuits have been difficult to verify in experiments.
Recently, a theory on synchronous activity propagation in a feedforward network has been demonstrated in vitro-silico \cite{NatNeuro:Reyes2003}, and has established one of the rare connections between theory and experiments in the field of neuroscience.

If identical neurons receive the same and temporally modulated input,
the spike timing would be roughly synchronous even if the initial membrane potentials were distributed by noise \cite{Science:Mainen1995}. The question in a feedforward network is, whether the timing of spikes within a layer becomes more synchronized or not as activity propagates through a sequence of neural layers.
When feedforward connection is uniform with excitatory efficacy as a whole, 
 the synchronous spike packet propagation is stable \cite{Abeles1991,Diesmann99,Cateau2001,Kistler2002,JNeurosci:Litvak2003}. 
This model is called the synfire chain \cite{Abeles1991}, and experimental evidence for this model has been reported {\it in vivo} \cite{JNeurophysiol:Abeles93}, and both {\it in vivo} and {\it vitro} \cite{Science:Ikegaya2004}.

The connection in the cortex is, however, not uniform but inhomogeneous, 
and composed of excitatory and inhibitory synapses. 
It would be more reasonable for a feedforward network to have such inhomogeneous connectivity.
For example, Mexican-Hat type connectivity, which excites nearby neurons and inhibits distal surrounding neurons, is a prevalent structure in the cortex.
As an inhomogeneous structure generates inhomogeneous input to the next neural layer,
it is not clear what types of spatio-temporal spike patterns are stable, and whether the spike packets are synchronized or not.
In this letter, we study the dynamics of a feedforward network with Mexican-Hat type connectivity to explore the effect of more biologically plausible connection on the propagation stability of spike packets, and the dependence of propagation speed on its own spatio-temporal activity pattern.
\begin{figure}
\begin{center}
\includegraphics{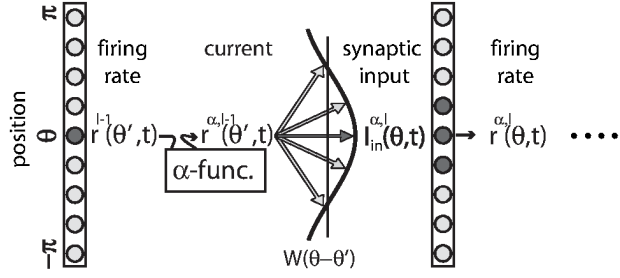}
\end{center}
  \caption{Network architecture. Each layer consists of $N$ units of neuron, which is arranged in a circle. Each neuron projects their axons to the post-synaptic layer with Mexican-Hat type connectivity.}
\label{fig:FMHModel}
\end{figure}

Model.-- Consider a structured feedforward network composed of identical Leaky Integrate-and-Fire (LIF) neurons. Each neuron is aligned in a circled neural layer, and they project axons to the next neural layer with the Mexican-Hat type connectivity (Fig. \ref{fig:FMHModel}). 
The input to one neuron includes output from pre-synaptic neural layer and random noisy synaptic current.
The dynamics of the membrane potential $v^l_\theta$ of a neuron at position $\theta$ on layer $l$ can be approximated into a stochastic differential equation,
\begin{align}
C \frac{\d v^l_\theta}{\d t} &= -\frac{v^l_\theta}{R} +  I^{\alpha,l}_{\mathrm{in}}(\theta,t) + \tilde{\mu}+ D' \eta(t), \label{eq:1}\\
I^{\alpha,l}_{\mathrm{in}}(\theta,t)&= \int^\pi_{-\pi}\frac{\d \theta'}{2\pi} W(\theta-\theta')r^{\alpha,l-1}(\theta',t), \label{eq:4}\\
r^{\alpha,l-1}(\theta,t) &= \int^0_{-\infty} \d t' \alpha(t')r^{l-1}(\theta,t-t'), \label{eq:5}\\
W(\theta) &= W_0 + W_1 \cos(\theta),
\end{align}
\begin{figure}
\begin{center}
\includegraphics{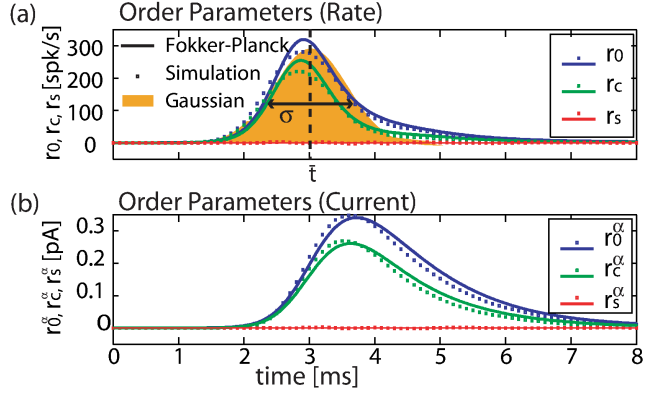}
\end{center}
  \caption{(a): Time courses of $(r_0(t), r_c(t), r_s(t))$, and the Gaussian approximation of $r_0(t)$ are shown. The Gaussian distribution has mean value of $\bar{t}$ and the standard deviation of $\sigma$. 
The approximated Gaussian curve is obtained by minimizing the mean squared error with $r_0(t)$ and the Gaussian curve. 
(b): The time courses of $(r^\alpha_0(t), r^\alpha_c(t), r^\alpha_s(t))$ are exhibited which correspond to the order parameter in the dimension of current. 
In both figures, results from numerical simulations with $10^4$ LIF neurons (squares) 
and the Fokker-Planck equation (solid lines) are shown.}
  \label{fig:OutputPDEODE}
\end{figure}

where $C$ is the membrane capacitance, $R$ is the membrane resistance, $\tilde{\mu}$ is the mean of noisy input, $\eta(t)$ is a Gaussian random variable satisfying $\langle\eta(t)\rangle = 0$ and $\langle \eta(t)\eta(t') \rangle = \delta(t-t')$. $D'$ is the amplitude of Gaussian noise. 
An input current $I^{\alpha,l}_{\mathrm{in}}(\theta,t)$ is obtained by the weighted sum of synaptic currents $r^{\alpha,l-1}(\theta,t)$ generated by pre-synaptic neurons. 
The synaptic current is derived from the convolution of firing rate $r^{l-1}(\theta,t)$ with the post-synaptic current time course $\alpha(t)$. Here, $\alpha(t) = \beta \alpha^2 t \exp(-\alpha t)$ where $\beta$ is chosen such that 
single EPSP generates $0.0014$ mV depolarization from resting potential.
The Mexican-Hat type connectivity consists of a uniform term $W_0$ and a spatial modulation term $W_1\cos(\theta)$ \cite{PNAS:BenYishai95}. The membrane potential dynamics follows the spike-and-reset rule: When $v^l_\theta$ reaches the threshold $V_{\mathrm{th}}$, a spike is fired and $v^l_\theta$ is reset to the resting potential $V_{\mathrm{rest}}$.
Throughout this paper, the parameter values are fixed as follows: $C = 100$ pF, 
$R = 100$ M$\Omega$, 
$V_{\mathrm{th}} = 15$ mV,
$V_{\mathrm{rest}} = 0$ mV,
$D' = 100$,
$\tilde{\mu} = 0.075$ pA,
$\alpha= 2$, and
$\beta = 0.00017$. 
\begin{figure*}[t]
\begin{center}
\resizebox{130mm}{!}{\includegraphics{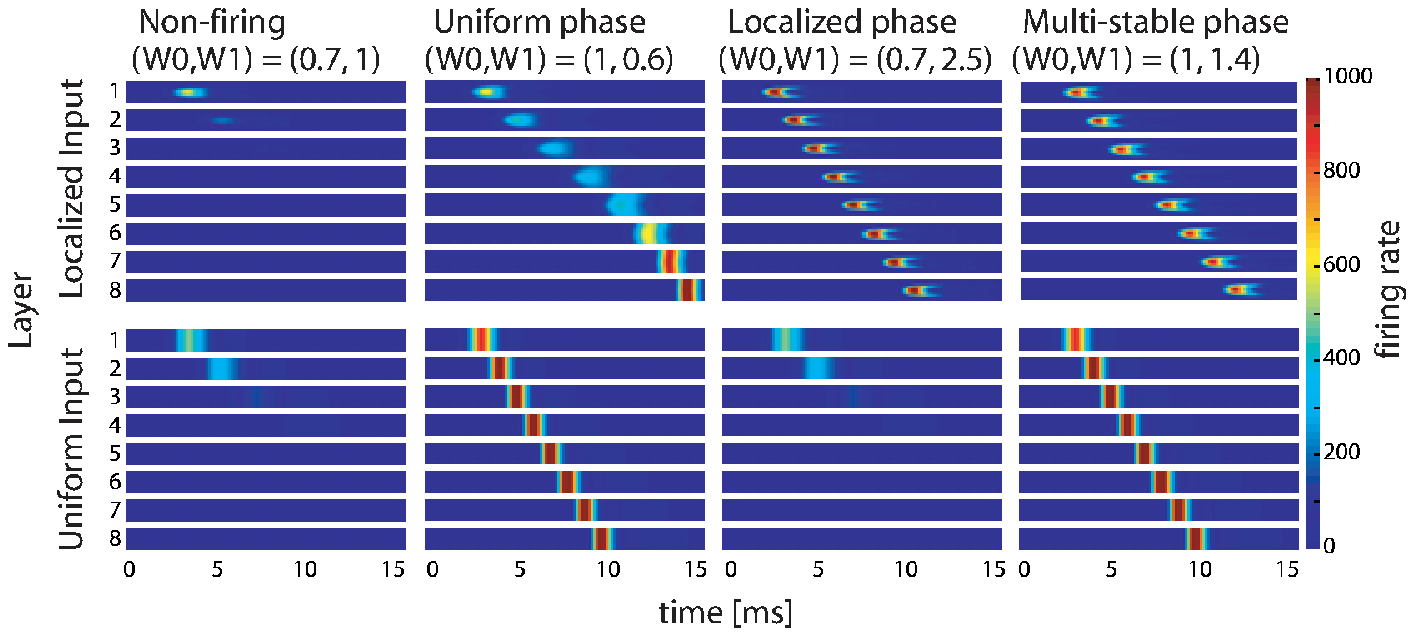}}
\end{center}
  \caption{Activities of 8 layers of feedforward network with four characteristic parameter sets. Evolutions of firing rate changes are illustrated with colors (see right color-bar). Figures on the upper figures are the response to an identical localized pulse input, and the lower figures are those of an uniform pulse input. Figures on the same column have the same Mexican-Hat type connectivity. There are four phase where different combination of spike packets are stable.}
\label{fig:All_MH}
\end{figure*}

Theory.-- The prerequisites for the full description of the activity of the network are time series of order parameters at arbitrary time $t$ defined as follows:
\begin{align}
r^l_0(t) &= \int^\pi_{-\pi} \frac{\d \theta}{2\pi} r^{l}(\theta,t), \label{eq:6}\\
r^l_c(t) &= \int^\pi_{-\pi} \frac{\d \theta}{2\pi} r^{l}(\theta,t)\cos(\theta), \label{eq:7}\\
r^l_s(t) &= \int^\pi_{-\pi} \frac{\d \theta}{2\pi} r^{l}(\theta,t)\sin(\theta), \label{eq:8}
\end{align}
where $r^l_0(t)$ is a population firing rate of the neuron population, and $r^l_c(t)$ and $r^l_s(t)$ are the coefficients of the Fourier transformation of the spatial firing pattern, 
which represent the spatial eccentricity of activity at time $t$.
Input currents are described with the order parameters as follows:
\begin{align}
 I^{\alpha,l}_{\mathrm{in}}(\theta,t) &= W_0 r^{\alpha,l-1}_0 (t)  \notag \\
& + W_1 \left( r^{\alpha,l-1}_c (t)\cos(\theta)  + r^{\alpha,l-1}_s (t)\sin(\theta) \right),
\end{align}
where $r^{\alpha,l}_{\{0,c,s\}}(t) = \int^0_{-\infty} \d t' \alpha(t') ~r^l_{\{0,c,s\}}(t-t')$ are also order parameters with the dimension of current.
Given the time sequence of order parameters in pre-synaptic layer, the order parameters in post-synaptic layer are obtained through the following calculations.
The stochastic equation (Eq. \ref{eq:1}) is equivalent to the Fokker-Planck equation \cite{Cateau2001,Book:Risken1996} in the limit of neuron number $N \rightarrow \infty$. Therefore, the probability density of membrane potential, $P^l_\theta(v,t)$, of neuron $\theta$ on layer $l$ at time $t$ obeys
\begin{equation}
\partial_t P^l_\theta = \partial_v \left( \frac{v}{\tau} - \frac{I^{\alpha,l}_{\mathrm{in}}(\theta,t)+\tilde{\mu}}{C} + \partial_v D \right)P^l_\theta,
\end{equation}
where $D = \frac{1}{2}\left( \frac{D'}{C}\right)^2$, and $\tau = RC$. 
One neural layer is divided into $100$ regions to calculate the Fokker-Planck eqs.
The resetting mechanism of LIF neuron requires absorbing boundary condition at threshold potential,$P^l_\theta|_{v = V_{\mathrm{th}}} = 0,$, and the current source at resetting potential gives, 
\begin{align}
r^{l}(\theta,t) & = \partial_{v}P^l_\theta \vert_{v = V^{+}_{\mathrm{rest}}}-\partial_{v}P^l_\theta \vert_{v = V^{-}_{\mathrm{rest}}},  \label{eq:CurrSource}\\
= & \left( \frac{V_{\mathrm{th}}}{\tau} - \frac{I^{\alpha,l}_{\mathrm{in}}(\theta,t)+\tilde{\mu}}{C} + {\partial_v}D \right)P^l_\theta|_{v = V_{\mathrm{th}}} \label{eq:Rate}.
\end{align}
The time series of pre-synaptic layer order parameters, $(r^{l-1}_0(t'), r^{l-1}_c(t'), r^{l-1}_s(t'))$ in the range of  $t'< t$ and the last condition (Eq. \ref{eq:Rate}) gives the firing rate of each neuron. 
Applying the definitions of the order parameters in Eqs.(\ref{eq:6})-(\ref{eq:8}) again, we can obtain the time series of order parameters on the post-synaptic neural layer $(r^l_0(t), r^l_c(t), r^l_s(t))$.
This sequential process of calculating the order parameters is essentially equivalent to the former analysis \cite{Neurocomputing:Hamaguchi2005} which is restricted to a binary neuron model with no memory (McCulloch-Pitts neuron model). The difference of the analysis between LIF neuron model and the McCulloch-Pitts neuron model is that the former one requires order parameters function depending on time, and the latter uses only the order parameters on time step before.

Results.-- We show an example of the time courses of $r^l_0(t), r^l_c(t), r^l_s(t)$ in response to a synchronized input in Fig. \ref{fig:OutputPDEODE} for both the numerical simulation of LIF neurons and the Fokker-Planck equation. 
The time course of order parameters calculated from the Fokker-Planck formulation (solid lines) well matches those of LIF neurons (squares).  
\begin{figure}
\begin{center}
\includegraphics{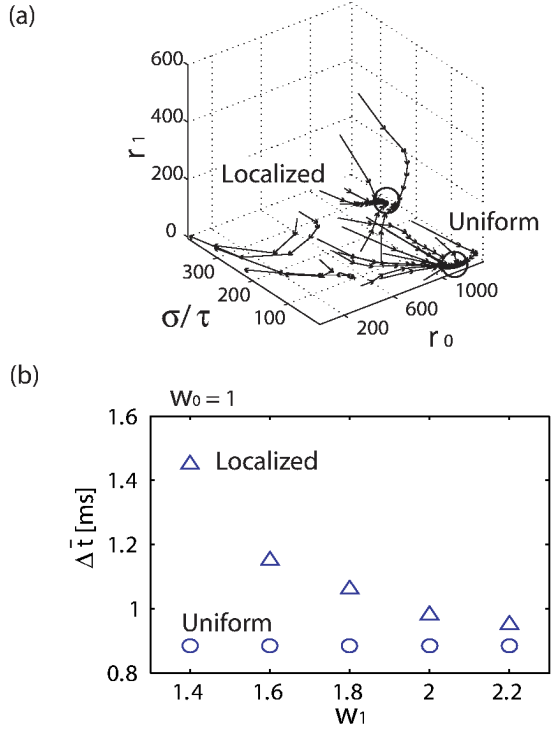}
\end{center}
  \caption{(a): Flow diagram with parameter $(W_0, W_1) = (1, 1.5)$ where both localized and uniform spike packets are stable. 
Two attractors in high $r_1$ region (Localized) and high $r_0$ with $r_1 = 0$ region (Uniform) are shown. 
A sequence of arrows indicates the evolution of a spike packet in the $(r_0,\sigma/\tau,r_1)$ space. (b): Plot of propagation time $\Delta \bar{t}$ that is necessary time for a spike packet to propagate one neural layer. This results indicate that localized spike packets propagate slower than the uniform ones. }
  \label{fig:PropSpeed}
\end{figure}
Our observation of activities propagating through the network with various parameter sets indicate that
there are two types of spike packets that stably travels along the network.
In Fig. \ref{fig:All_MH}, we can see the responses of this network with four characteristic $(W_0, W_1)$ parameter sets. 
Each figures in the same row shares the same input time sequence.
Here, input to initial layer is formulated in terms of the firing rate on the pre-synaptic layer,
$r^{l}(\theta,t) = \frac{r_0 + r_1 \cos(\theta)}{\sqrt{2\pi \sigma^2}} \exp(-\frac{(t-\bar{t})^2}{2 \sigma^2}).$
Here we use $r_0 = 500, r_1 = 350$ for the upper figures as a localized input and $r_0 = 900, r_1 = 0$ for the lower ones as a uniform input. The common parameters are $\sigma = 1$ and $\bar{t} = 2$.
When both $W_0$ and $W_1$ are small, no spike packet can propagate (Non-firing). 
When the uniform activation term $W_0$ is sufficiently strong, 
a uniform spike packet is stable (Uniform phase). 
Note that even if a localized input elicits a localized spike packet with several layers, it finally decays to the uniform spike packet (Fig. \ref{fig:All_MH}, upper figure of Uniform phase).
When the Mexican-Hat term $W_1$ is strong enough, 
only a localized spike packet is stable (Localized phase). 
This local synchronous activity propagation is characteristic of this network and we call this a localized synfire chain.
When $W_0$ and $W_1$ are balanced with a certain ratio, there exists a novel firing mode where both uniform and localized spike packets are stable depending on the initial input (Multi-stable Phase). 

For the quantitative evaluation of the shape of a spike packet, 
we define indices $r_0, r_1$, and $\sigma$ to characterize the spatio-temporal pattern of a spike packet.
$(r_0, r_1)$ can be directly defined as follows:
\begin{align}
r_0 &= \int \d t ~r_0(t) - \nu_{\mathrm{spont}}, \\
r_1 &= \int \d t ~r_1(t) ,~~ r_1(t) = \sqrt{(r_c(t))^2 + (r_s(t))^2}.
\end{align}
These values are a natural extension of indices used in a study on the synfire chain \cite{Diesmann99}.
The basic idea of characterizing the spike packet was to approximate the firing rate curve with a Gaussian function as shown in Fig. \ref{fig:OutputPDEODE} (yellow region), 
and the area and the variance of the Gaussian curve were used as indices of spike packets. 
Instead of these, we directly use the area of firing rate $r_0(t)$ minus the spontaneous firing rate $\nu_{\mathrm{spont}}$ because the area of Gaussian is just an approximation of the area of $r_0(t)$.
Further, since we have another order parameter $r_1(t)$, we can naturally derive a new index $r_1$
as the area of order parameter $r_1(t)$. This index indicates the eccentricity of firing patterns.
The index $\sigma$, representing the measure of synchrony, is obtained as the variance of the Gaussian curve(Fig. \ref{fig:OutputPDEODE}). 
To analyze the propagation speed of spike packets, the index of mean spike timing  $\bar{t}$ is taken from the peak time of the Gaussian curve. 

In the Uniform phase, input with an intensity more than certain threshold generates a uniform spike packet.
In this phase, this system can classify the input intensity into two states because there are two stable attractor; a uniform spike packet and a silent state.
In the Localized phase, it has two attractors, a localized spike packet and a silent state, and information of the input position can be encoded. 
This network, however, cannot discriminate between no-input and strong but spatially uniform input. Because a uniform activity is unstable in this phase, and it leads to a non-firing state.
In contrast, the Multi-stable phase has three attractors. The flow diagram illustrated in Fig. \ref{fig:PropSpeed}(a) shows the evolution of spike packets in the $(r_0,\sigma/\tau,r_1)$ space. It shows the existence of three attractor, uniform and localized spike packet, and non-firing state (region where $\sigma$ is very large). The Multi-stable phase can encode the above difference, and may play a functional role in enhancing the ability to discriminate the input difference. 

The propagation time for a spike packet obviously depends on the connection $W_0$ and $W_1$.
Furthermore, spike packets in the Multi-stable phase in Fig. \ref{fig:All_MH} 
indicate that the speed also depends on the propagating spike pattern. 
To investigate this effect, we calculate the difference of propagation times  $\Delta\bar{t} = \bar{t}_{\mathrm{post}}- \bar{t}_{\mathrm{pre}}$  with various $W_1$ parameters after spike packets reach their stable states (Fig. \ref{fig:PropSpeed}(b)). 
Within the plotted region, higher $W_1$ reduces the propagation time of a localized spike packet. In contrast, the propagation time of the uniform spike packet does not depend on $W_1$ because uniform activity leads to $r_c(t) = r_s(t) = 0 $, so $W_1$ term can be neglected.
These results suggest that the speed of information processing in the brain depends on spiking patterns, or the representation of information.

The model of a localized synfire chain provides a common test bed for several fields of neuroscience.
It is not only theoretically tractable but also one of a few models that is experimentally verifiable \cite{NatNeuro:Reyes2003}.
The localized synfire chain also has strong connection to population coding \cite{NeuralComp:Wu2002,NatRevNeurosci:Pouget2000,AnnuRevNeurosci:Pouget2003}, because many of their neural substrate is also the Mexican-Hat type connectivity.
Therefore, we believe that our results can bridge theoretical, computational, and experimental neuroscience and provide a deeper understanding of brain functions.

This work is partially supported by 
the Advanced and InnovationalResearch Program in Life Sciences,
a Grant-in-Aid Scientific Research, on Priority Areas (2) No. 15016023 Advanced Brain Science Project, 
Priority Areas No.14084212, Scientific Research (C) No.16500093.
from the Ministry of Education, Culture, Sports, Science, and Technology, the Japanese Government. 


\bibliography{../../../Bibtex/article,../../../Bibtex/books}

\end{document}